\documentclass[twocolumn,dvipsnames]{aastex631}
\usepackage[T1]{fontenc}
\usepackage{amsmath}
\usepackage{natbib}
\usepackage{soul}
\usepackage{xcolor}
\usepackage{lipsum}
\usepackage{float}
\usepackage{graphicx}	
\usepackage{multirow,makecell}
\usepackage{placeins}
\usepackage{bm,amsfonts,hhline,amssymb,microtype,float,latexsym,epsf,mathtools,times,makecell,array}

\usepackage{tabularx}
\usepackage{epstopdf}
\usepackage{gensymb}
\usepackage{pifont}
\usepackage{hyperref}
\hypersetup{colorlinks=true, linkcolor=red, filecolor=magenta, urlcolor=Sepia, citecolor=blue}

\definecolor{ForestGreen}{HTML}{228B22}

\received{XXX}
\revised{YYY}
\accepted{ZZZ}

\shorttitle{Gravitational wave from SSSs}
\shortauthors{Das, Bulik, Roy \& Mukhopadhyay}

\begin{document}



\title{Continuous Gravitational Waves from Supersoft X-ray Sources: Promising Targets for deci-Hz Detectors} 

\author[0009-0004-6340-9337]{Mayusree Das$^{*}$}
\thanks{mayusreedas@iisc.ac.in}
\affiliation{Joint Astronomy Programme, Department of Physics, Indian Institute of Science, Bangalore, 560012, India}

\author[0000-0003-2045-4803]{Tomasz Bulik$^{\dagger}$}
\thanks{tb@astrouw.edu.pl}
\affiliation{Astronomical Observatory, University of Warsaw, Al. Ujazdowskie 4, 00478 Warszawa, Poland}

\author[0009-0005-4792-0378]{Sreeta Roy$^{\S}$}
\affiliation{Astronomical Observatory, University of Warsaw, Al. Ujazdowskie 4, 00478 Warszawa, Poland}
\thanks{sroy@astrouw.edu.pl}

\author[0000-0002-3020-9513]{Banibrata Mukhopadhyay$^{\ddagger}$}
\thanks{bm@iisc.ac.in}
\affiliation{Department of Physics, Indian Institute of Science, Bangalore, 560012, India}
\affiliation{Joint Astronomy Programme, Department of Physics, Indian Institute of Science, Bangalore, 560012, India}

\begin{abstract}
Supersoft X-ray sources (SSSs) host white dwarfs (WDs) accreting at rates that sustain steady nuclear burning, driving rapid mass growth, radial contraction, and magnetic field amplification. Angular-momentum transfer from the accretion disk naturally spins up the WD, while the amplified internal magnetic field induces a non-axisymmetric deformation in presence of a misaligned rotation. Such WDs emits continuous gravitational waves (CGWs). We model the coupled evolutions of stellar mass, spin, and magnetic structure in accreting WDs in SSSs with \texttt{MESA}, and compute the resulting quadrupolar deformation with the Einstein-Maxwell solver \texttt{XNS}. We show that WDs in SSSs, particularly near the end ofthermal timescale mass transfer and close to the Chandrasekhar mass limit, produce CGWs predominantly in the deci-Hz band accessible to planned detectors such as DECIGO, BBO, Deci-Hz, ALIA, and LGWA, and are distinguishable from other Galactic CGW sources such as AM\,CVn systems, detached double WDs, and isolated WDs. Well-studied SSSs such as CAL~83 and RX~J0019+2156 can be detectable, enabling targeted CGW measurements that directly probe WD's internal magnetic fields and rotation, while blind searches can reveal hundreds of obscured SSSs otherwise missed in soft X-rays and map the hidden population of accreting, rapidly rotating, magnetized WDs in nearby galaxies. A CGW detection from WDs in SSSs could also identify potential pre-explosion Type~Ia progenitors.
\end{abstract}

\section{Introduction}

Supersoft X-ray sources (SSSs) were first identified by ROSAT with luminosity $\sim10^{37}$--$10^{38}\,\mathrm{erg\,s^{-1}}$ and effective temperature of $10^5$K ($kT\!\sim\!20$--$50$ eV; \citealt{Trumper1991,Greiner1991}). Their optical counterparts confirm that they are close binaries with orbital periods of $\sim$8\,hr--1.4\,d. Later they were understood as interacting binaries where a white dwarf (WD) accretes H-rich material at $\dot M_{\rm WD}\!\sim\!(1$--$5)\times10^{-7}\,M_\odot\,\mathrm{yr^{-1}}$ from a slightly evolved main sequence star (MS), subgiant, or red-giant donor, sustaining stable H burning \citep{van-den-Heuvel1992,Kahabka1995,Nomoto2007,Wolf2013}. A high mass transfer rate from the donor ($\dot M_{2}$) arises during thermally unstable Roche-lobe overflow.
When $\dot M_{2}$ exceeds the WD’s critical burning rate (set by the Eddington luminosity), $\sim$60--80\% of the transferred mass is expelled in an optically thick wind, so that only $\sim$20--40\% is retained by the WD, defining the actual accretion rate $\dot M_{\rm WD}$ \citep{HachisuKato2003a}.

The supersoft X-ray luminosity is generated by energy released from steady nuclear burning on the WD surface rather than by the gravitational potential energy liberated by accretion, although absorption by circumstellar material associated with wind losses can modify the observed emission. At lower $\dot M_{2}$, nuclear burning becomes unstable, leading to recurrent novae with negligible mass retention, while at higher $\dot M_{2}$ the burning envelope expands toward a red-giant like configuration that again drives strong winds \citep{Nomoto1979}.

\begin{figure}
\begin{center}
\includegraphics[width=0.75\columnwidth]{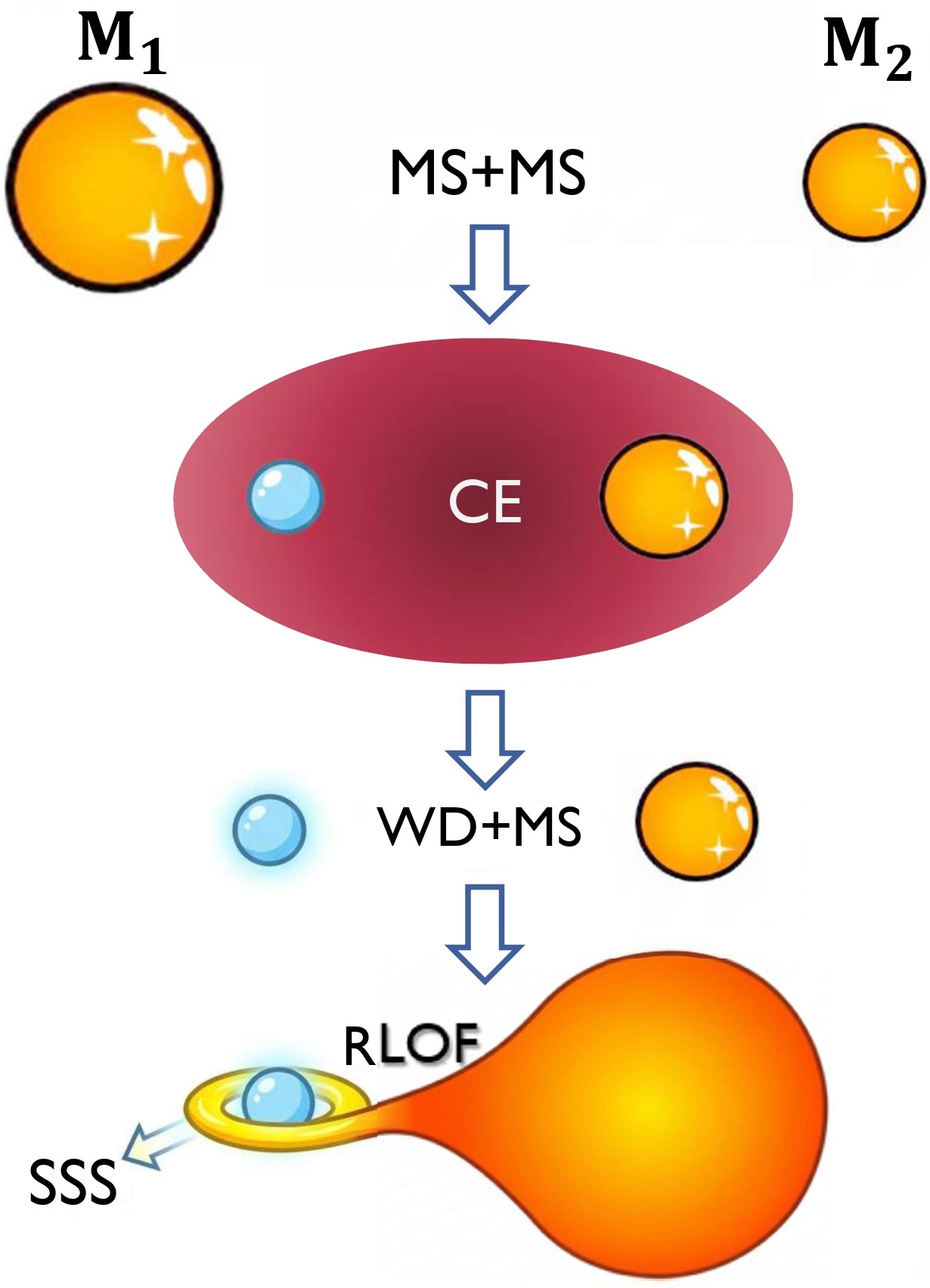}
\caption{Cartoon illustration of the standard evolutionary channel leading to a SSS.}
\label{fig:evolcartoon}
\end{center}
\end{figure}

SSSs naturally emerge from $\sim$5--8\,$M_\odot$ MS (primary) and $\sim$1--3\,$M_\odot$ MS (secondary; $M_2$) binary evolution \citep{Rappaport1994,Hachisu1999}. After an initial common-envelope episode, the primary becomes a CO/ONe WD in a compact orbit with secondary MS. After $\sim 8/(M_2/M_\odot)^2$ Gyr, the secondary evolves to fill its Roche lobe, and transfers mass on a thermal timescale onto the WD, producing the SSS phase (see Figure \ref{fig:evolcartoon}). If the WD is initially massive, stable accretion can grow it toward the Chandrasekhar limit, making SSSs plausible progenitors of $\sim$10--30\% of Type~Ia Supernovae \citep{Rappaport1994,Hachisu1999}. Otherwise, after thermal-timescale ($\sim3\times10^{7}(M_2/M_\odot)^{-2.3}$ yr), $\dot M_2$ drops below the steady-burning regime, after which the system transitions into late-SSS states such as V~Sge--like systems with accretion--wind cycle, symbiotic SSSs, or recurrent novae (e.g., U~Sco) or eventually evolves into a double WD.

Population synthesis models predict existence of $\gtrsim1000$ SSSs in the Milky Way (MW) \citep{Rappaport1994}, but due to strong interstellar absorption in the Galactic plane only $\sim$10 are observed \citep{van-den-Heuvel1992}. In the Large Magellanic Cloud (LMC), in which direction interstellar column densities are lower, $\sim$30 SSSs are detected out of several hundreds \citep{Rappaport1994}. CAL~83 is the best studied prototypical SSS, alongside nearby Galactic systems such as RX~J0019+2156 and RX~J0925-4758 \citep{Greiner2000}, however, the latter shows dense winds and transient jets consistent with intermittent accretion and wind regulated burning \citep{Motch1994}.

Steady accretion in SSSs increases the WD mass and leads to contraction of the stellar radius, amplifying the magnetic field through flux freezing, while angular momentum exchange with the accretion disk spins up the star \citep{GhoshLamb1979b}. In SSSs, steady accretion requires that the magnetic field remains below the propeller threshold \citep{Ghosh1995}, so that the ram pressure of the inflowing matter exceeds the magnetic pressure and accretion onto the WD surface can proceed. Magnetic field amplification and spin-up therefore occur within this sub-propeller regime during the SSS phase, prior to any possible transition to propeller-dominated states at later evolutionary stages.

In such magnetized, spinning WDs residing in SSSs, anisotropic magnetic pressure induces a quadrupolar deformation, and a misalignment between the magnetic and spin axes (obliquity angle $\chi$) produces continuous gravitational waves (CGWs) at the spin frequency $\nu$ and its harmonic $2\nu$ \citep{BG1996,heyl2000}. Observed magnetic dichroism in several WDs \citep{Putney1992,Suto2023} confirms that oblique magnetic geometries ($\chi\!\neq\!0$) are common. Thus SSSs are favorable gravitational wave (GW) targets: continuous high accretion naturally couples WD mass growth, magnetic field amplification, and rotational spin-up -- three key ingredients for persistent CGW emission. The CGW can be enhanced and thus highly plausible for detection near the end of the thermal timescale mass transfer phase or when the WD approaches the Chandrasekhar limit. Such signals may fall in the sensitivity windows and frequency range of planned/proposed GW missions including LISA \citep{LISAdefinitionstudy}, DECIGO \citep{Kawamura2020}, the proposed Indian Deci-Hz detector (Rajesh K. Nayak et al., private communication)\footnote{Based on representative noise PSD of a possible Indian deciHz mission, tentatively referred to as IndIGO-D.}, ALIA \citep{Crowder2005}, BBO \citep{CutlerHarms2006}, TianQin \citep{TianQinoverview}, LILA \citep{Creighton2025_LILA}, and LGWA \citep{Ajith2024_LGWA}. Apart from CGW from the WD, SSSs also emit GW due to binary orbital evolution, which lies below the frequency range considered here; such orbital GW emission from accreting double WD binaries has been studied by \citet{Yi2024}.

Detection of SSSs such as CAL~83 or RX~J0019+2156 in CGWs could probe the WD’s internal magnetic field and angular velocity. Moreover, blind searches could reveal large hidden population of SSSs in the MW and LMC that remain invisible in soft X-rays otherwise due to interstellar absorption. Finally, if near-Chandrasekhar SSSs--plausible Type~Ia progenitors, such as CAL~83 (hosting a $\sim$1.3\,$M_\odot$ WD) are detected as CGW sources, they would provide a direct identification of a pre-explosion Type~Ia progenitor. Type~Ia supernova yet has not been observed to erupt from a system previously confirmed as a progenitor, so a  pre-explosion CGW identification would offer a predictive path toward an eventual optical supernova discovery.

In this work, we link the evolution of accreting, magnetized, rapidly rotating WDs in SSSs to their CGW emission and assess detectability for both known systems and blind searches, extending our isolated WD study \citep{Das2025_MWD_CGW}; hereafter DMB2025). Section \ref{sec:wdmodel} outlines our stellar evolution models for the accreting WDs. Section \ref{sec:gwmodel} presents the magnetic deformation and CGW signal computed with the stellar structure solver and discusses their detectability with future detectors. We conclude by summarizing the implications of CGW detection for both known SSSs and the hidden SSS population.

\section{Modeling and evolution of Magnetized, Rotating, Accreting White Dwarfs}
\label{sec:wdmodel}

\subsection{Stellar Structure and Evolution with \texttt{MESA}}

To model stellar evolution of magnetized, rotating, and accreting WDs, we employ \texttt{MESA} \citep{Paxton11}. Thereafter, we first outline the 
stellar-structure and evolution equations involved in the standard (non-magnetic, non-rotating, non-accreting) configuration. These baseline equations form the reference framework to which rotation, magnetic fields, and accretion physics will be added in the following subsections. The stellar-structure equations in Lagrangian form \citep{Paxton11} are:
\begin{equation}
\frac{dr}{dm} = \frac{1}{4\pi r^2 \rho},
\label{eqdm}
\end{equation}
\begin{equation}
\frac{dP}{dm} = -\frac{Gm}{4\pi r^4},
\label{eqdP}
\end{equation}
\begin{equation}
\frac{dT}{dm} = -\frac{Gm}{4\pi r^4}\frac{T}{P}\nabla,
\label{eqdT}
\end{equation}
\begin{equation}
\frac{dL}{dm} = \epsilon_{\rm nucl}- \epsilon_{\nu,{\rm therm}}+ \epsilon_{\rm grav},
\label{eqnuc}
\end{equation}
\begin{equation}
\frac{dX}{dt}
= \left(\frac{dX}{dt}\right)_{\rm burn}
+ \frac{d}{dm}\left(\sigma \frac{dX}{dm}\right).
\label{eqX}
\end{equation}
Here $\nabla \equiv d\ln T/d\ln P$ is the temperature gradient, 
$\epsilon_{\rm nucl}$ is the nuclear energy generation rate per unit mass,
$\epsilon_{\nu,{\rm therm}}$ is the neutrino energy loss rate per unit mass from thermal processes, and
$\epsilon_{\rm grav}$ is the gravothermal energy generation rate per unit mass during compression or expansion (see \citealt{Paxton11}),
$(dX/dt)_{\rm burn}$ represents rate of changes in composition from nuclear burning, $\sigma$ is the Lagrangian diffusion coefficient for compositional transport, $\rho$ is the density, $r$ is the radial coordinate and $T$, $P$, $L$, and $X$ denote the temperature, pressure, luminosity, and compositional mass fraction, respectively. \texttt{MESA} solves the above equations in an adaptive, time-dependent Lagrangian framework with mass $m$ as the independent coordinate, using the \texttt{Skye} equation of state \citep{Jermyn21} for the degenerate electron gas, with smooth transitions across partial and full degeneracy.

A key limitation of our present model is that we suppress steady H burning by setting $\epsilon_{\rm nucl}=0$ and $(dX/dt)_{\rm burn}=0$ for numerical stability, even though real SSS systems do undergo stable H burning producing He (the net effect is incorporated in Section \ref{sec:acc}). This may slightly alter envelope's temperature stratification but does not significantly affect the WD’s radius, spin, or magnetic field, which we study here.

\subsection{Magnetic Field}
\label{sec:mag}

Having outlined the non-magnetic baseline structure, we now incorporate magnetic fields into the WD model. Because \texttt{MESA} does not solve Maxwell equations, the magnetic configuration must be supplied externally. We adopt the field profiles obtained from the Einstein--Maxwell solver \texttt{XNS} \citep{sold2021main}, modified for WDs in DMB2025 (this
solver will be described in detail in Section~\ref{sec:gwmodel}). Observational and theoretical studies indicate that WDs may host strong internal toroidal components that dominate over the internal poloidal field
\citep{Wickramasinghe2014}. Motivated by these stable configuration, we represent the internal toroidal field using the XNS-motivated prescription
\begin{equation}
B(r) = \mathcal{K}\,\rho\, r,
\label{eq:magprof}
\end{equation}
where $\mathcal{K}$ sets the normalization. This scaling is similar to flux freezing, i.e. $B\propto \rho r \propto M/r^{-2}$, and thus $B(r)$ enhances locally under contraction.

However, the magnetic field also decays through Ohmic and Hall dissipation \citep{HK1998,bhattacharya22}, i.e.,
\begin{equation}
\frac{dB(r)}{dt}
= - B(r)\left(\frac{1}{t_{\rm ohm}} + \frac{1}{t_{\rm Hall}}\right),
\label{eq:Bdecay}
\end{equation}
with characteristic timescales \citep{GR1992,Cumming2002}
\begin{equation}
t_{\rm ohm}
= \left(2.2\times10^{11}\,{\rm yr}\right)\,
  \rho_{c,9}^{1/3}\, R_{8}^{1/2}\,
  \left({\rho_{\rm avg}}/{\rho_c}\right),
\end{equation}
\begin{equation}
t_{\rm Hall}
= \left(1.7\times10^{11}\,{\rm yr}\right)\,
  l_{8}^{2}\, B_{m,12}^{-1}\, \rho_{c,9},
\end{equation}
where for any quantity $X_n \equiv X/10^n$ in CGS units, $B_m$ is the maximum magnetic field of a profile, $R$ is WD's radius, $\rho_{\rm avg}$ and $\rho_c$ are the average and central densities, respectively.

At each \texttt{MESA} timestep, we compute the magnetic profile $B(r)$ using Equation~(\ref{eq:magprof}) for the current stellar structure including contribution from magnetic pressure (see below). The field is then evolved over the timestep using Equation~(\ref{eq:Bdecay}), and the normalization $\mathcal{K}$ is updated accordingly, ensuring that the magnetic profile remains consistent with the changing WD structure while incorporating Ohmic and Hall dissipations.

\begin{figure}
\begin{center}
\includegraphics[width=\columnwidth]{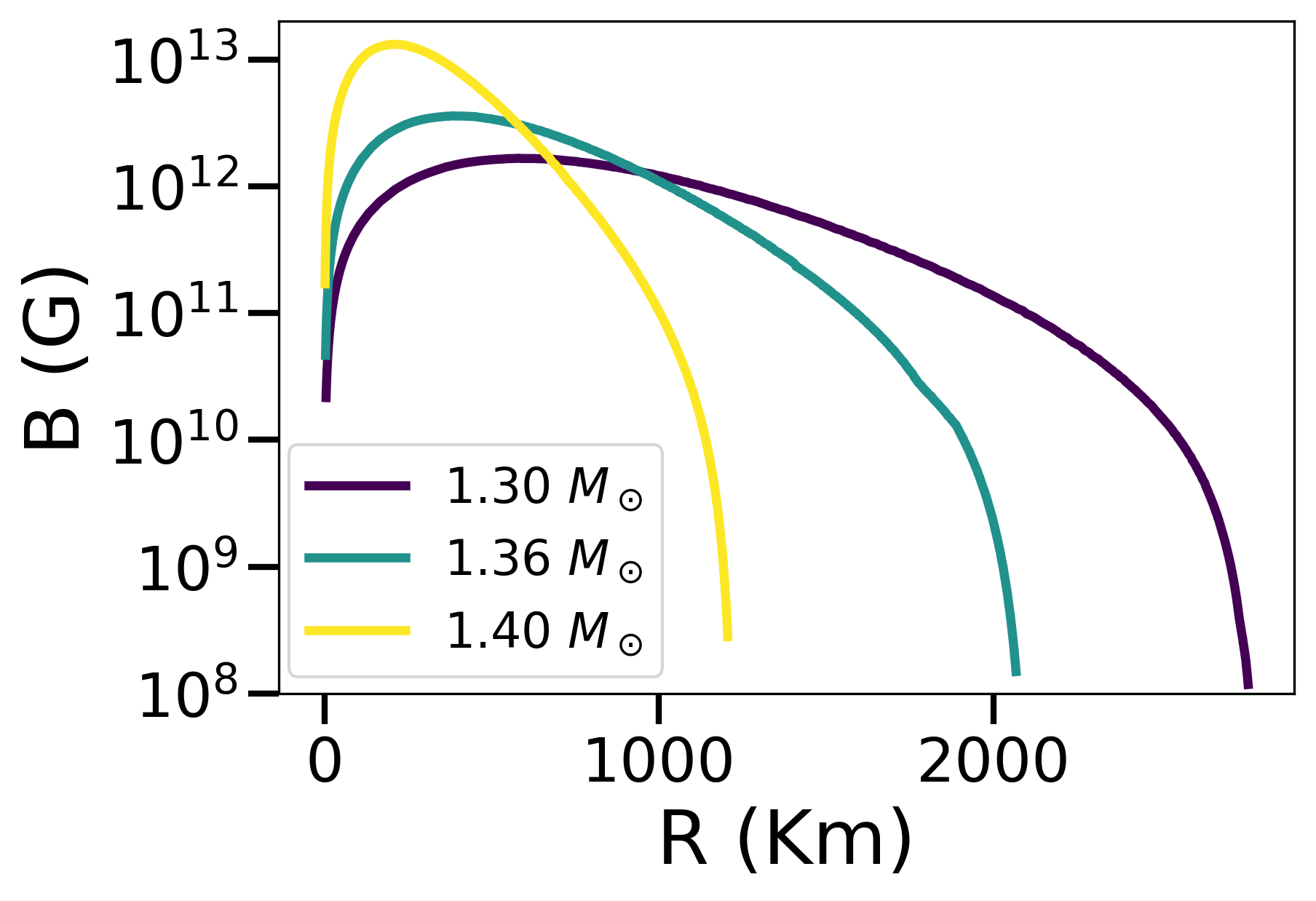}
\caption{Radial profiles of the internal toroidal magnetic field $B(r)$ at several
stages along the accreting WD sequence (Section~\ref{sec:acc}), showing the
amplification of $B$ due to mainly stellar contraction.}
\label{fig:magprof}
\end{center}
\end{figure}

An illustration of the magnetic field evolution for a WD of mass $1.3\,M_\odot$ that accretes and leads to $1.4\,M_\odot$ over $\sim 10^{6}$ yr (details of accretion is in Section~\ref{sec:acc}) is shown in Figure~\ref{fig:magprof}. As the star contracts during mass growth, an initial maximum internal field $B_{m}\sim10^{12}\,$G, interpreted as a fossil field inherited from a strongly magnetized progenitor Ap/Bp stars \citep{Ferrario2015, Quent2018,MukhopadhyayBhattacharya2022}, is amplified further to $\sim10^{13}\,$G. The corresponding magnetic to gravitational energy ratio increases from $7\times 10^{-7}$ to $4\times 10^{-4}$. Over the shorter accretion timescale considered here, Ohmic and Hall diffusions are negligible, since their decay times in the degenerate interior are $\gtrsim10^{8}$-$10^{9}$\,yr, so the evolution is governed predominantly by compression rather than dissipation.

The internal field modeled here is toroidally dominated, with the surface strength being aound four orders of magnitude weaker than $B_m$ (see Figure~\ref{fig:magprof}). In addition, a one to two order(s) of magnitude weaker large-scale poloidal field is expected to be present: its polar amplitude $B_p \approx B_m/10^{5-6}$ \citep{BraiSp2006,BR2009}. This exterior poloidal component is treated purely as an assumed dipole for radiation and disk-field coupling (to be discussed in Section \ref{sec:angmom}); it is not included in the internal equilibrium computed by \texttt{XNS} or \texttt{MESA} (as effect is negligible to impact on stellar structure).

This evolving (internal) field also feeds back on the stellar structure through its pressure support. The internal toroidal magnetic field contributes to an additional pressure term and thus the total pressure in Equation \eqref{eqdP} as
\begin{equation}
P = P_{\rm matter} + \frac{B^2}{8\pi},
\end{equation}
where $P_{\rm matter}$ is the thermodynamic matter pressure. The magnetic pressure contribution $B^2/8\pi$ is included in \texttt{MESA} through an \texttt{extra\_pressure} term implemented in \texttt{other\_pressure} routine of \texttt{run\_star\_extras}
\citep{Paxton19}, ensuring that the stellar structure remains consistent with the evolving magnetic field at each timestep.

\subsection{Rotation}
\label{sec:rot}

Having incorporated magnetic fields into the WD structure, we now include rotation, which modifies the stellar stratification. In \texttt{MESA}, rotation is treated using the shellular formalism of \citet{Endal1976}, where each isobar is replaced by an equivalent sphere of radius $r_p = (3V_p/4\pi)^{1/3}$, with $V_p$ the volume enclosed by the isobar. The continuity and energy conservation equations retain their usual form, while the hydrostatic equilibrium and energy transport equations acquire rotational correction factors $f_P$ and $f_T$, respectively, that account for the reduction of effective gravity and the geometric distortion of equipotential surfaces \citep{Paxton13}. The modified Equations \eqref{eqdP} and \eqref{eqdT} are therefore
\begin{equation}
\frac{dP_p}{dm_p} = -\,\frac{G m_p}{4\pi r_p^{4}}\,f_p, \qquad
\frac{dT_p}{dm_p} = -\,\frac{G m_p}{4\pi r_p^{4}}\,\frac{T_p}{P_p}\,\nabla\, f_T.
\end{equation}
Here $m_p$, $r_p$, $P_p$, and $T_p$ are quantities averaged over an isobar. $f_p$ and $f_T$ depend on the surface-averaged effective gravity $g_{\rm eff}=g-\Omega^2 r_\perp$, where $r_\perp$ is the cylindrical (equatorial) distance from the rotation axis, $\Omega$ is stars angular rotation rate. \texttt{MESA} \citep{Paxton13} adopts the analytical functions $f_p(\Omega)$, $f_T(\Omega)$ derived by \citet{Endal1976} in a one-dimensional (1D) form. In our models, $\Omega$ remains well below the critical (break-up) limit, with $\Omega/\Omega_{\rm crit}\lesssim 0.01$, so centrifugal distortion is minimal and $f_p\simeq f_T\simeq 1$, leaving the stellar structure close to non-rotating ones. Consistently, the ratio of rotational to gravitational energies remains small ($\lesssim 0.1$), confirming that none of the models approach rotational instability.


\subsection{Accretion}
\label{sec:acc}

Having incorporated both magnetic fields and rotation into the WD structure, we now include mass accretion, as occurs in SSSs. In our models, the companion is not explicitly evolved; instead, the net accretion rate onto the WD, $\dot M_{\rm WD}$, obtained after retention from the donor’s mass transfer rate $\dot M_2$, is prescribed following \citet{Paxton15}. WD's mass $M$ therefore increases as
\begin{equation}
M = M_\star + \dot M_{\rm WD}\,dt,
\label{eq:acc}
\end{equation}
where $M_\star$ is the initial mass. Newly accreted material is added to the outermost mesh cell, after which \texttt{MESA} readjusts the model to re-establish hydrostatic and thermal equilibrium. Consistent with SSS conditions, we treat the accreted material as pure He. Although the transferred gas is H-rich, stable H burning at the characteristic SSS accretion rates ($\dot M_{\rm WD}\sim10^{-7}\,M_\odot\,{\rm yr^{-1}}$) rapidly converts H to He \citep{Nomoto2007,Wolf2013}, on timescales much shorter than any structural timescale of the WD. As a result, the surface composition becomes effectively He-dominated, thus, treating the accreted material as pure He is a good approximation.

\begin{figure*}
\begin{center}
\includegraphics[scale=0.66]{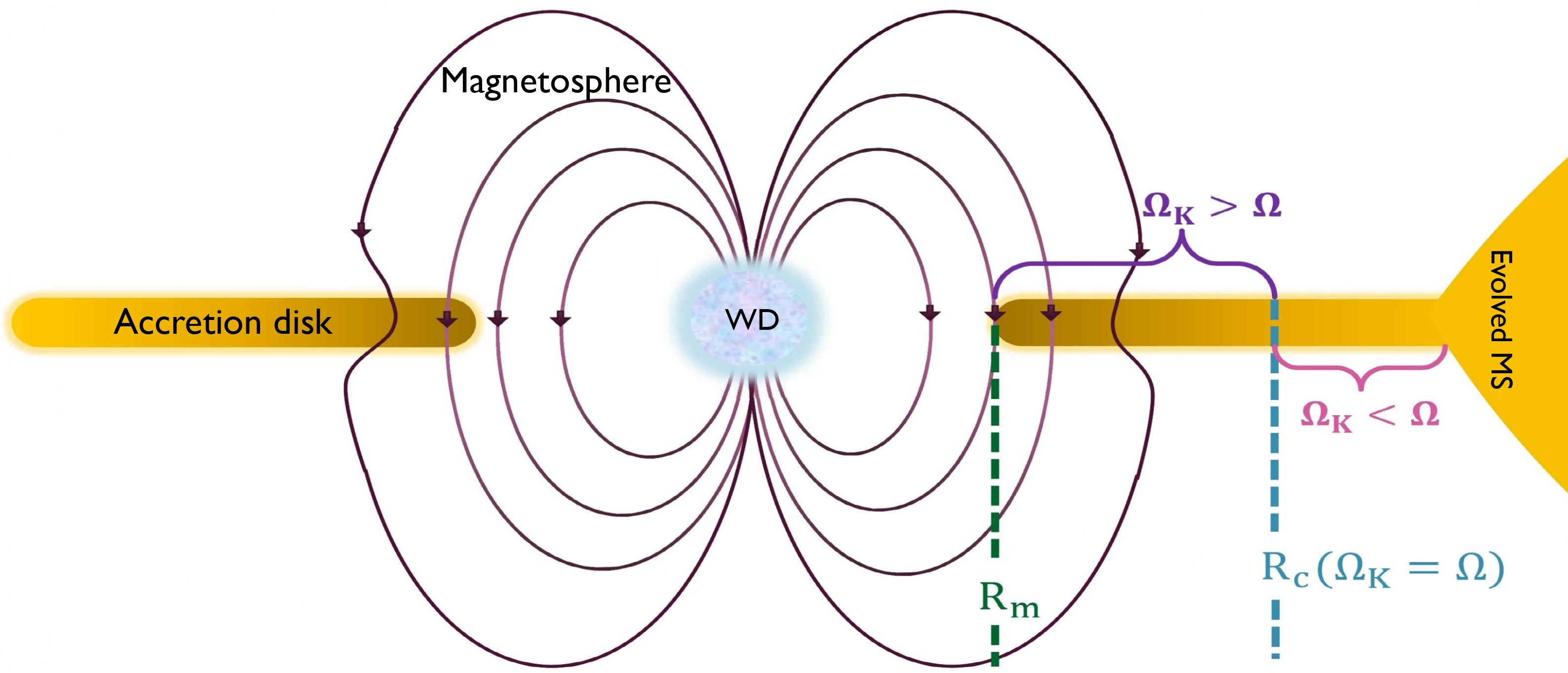}
\caption{Cartoon illustration from side view of the accretion disk and magnetosphere. See text for notations.}
\label{fig:glcatoon}
\end{center}
\end{figure*}

Matter transferred through Roche-lobe circularizes via angular-momentum conservation and forms a Keplerian accretion disk. Accretion proceeds as long as the inflowing matter can penetrate the stellar magnetosphere set by external dipolar field. The field then truncates the disk at a radius $R_{\rm m}$, where magnetic pressure balances the inflow ram pressure; inside $R_{\rm m}$ the disk is disrupted and accretion proceeds along magnetic funnels, as described in the classical Ghosh-Lamb (GL) model \citep{GhoshLamb1979a,GhoshLamb1979b}. This is shown as illustration in Figure \ref{fig:glcatoon}. Although originally derived for neutron stars, this model is also applicable for WDs \citep{Norton2004}. $R_{\rm m}$ is thus obtained as,
\begin{equation}
    \frac{B^2(R_{\rm m})}{8\pi} = \frac{1}{2}\rho_{\rm acc} v^2,
\end{equation}
where outside the star $B(r)=\mu/r^3$ with dipole moment 
$\mu = B_p R^3$,
$v = (2GM/R_{\rm m})^{1/2}$ is the free-fall speed,
$R$ is the WD radius, and 
$\rho_{\rm acc} \simeq \dot M_{\rm WD}/(4\pi R_{\rm m}^2 v)$.
Substituting these relations yields the standard expression for the \textit{magnetospheric radius}
\begin{equation}
    R_{\rm m} = \xi 
    \left( \frac{\mu^4}{2GM \dot M_{\rm WD}^2} \right)^{1/7},
\end{equation}
with $\xi \simeq 0.52$ for disk accretion \citep{GhoshLamb1979b}.

The Keplerian disk carries the specific angular
momentum of the binary orbit whose
angular velocity obeys $\Omega_{\rm K}(r) = \sqrt{(GM/r^3)}$  \citep{ShakuraSunyaev1973}. The inner disk generally rotates faster than the WD
\citep{GhoshLamb1979a}, while the outer disk rotates more slowly (see Figure \ref{fig:glcatoon}). The radius where the disk’s Keplerian frequency equals the stellar rotation rate is the \textit{corotation radius},
\begin{equation}
 R_{\rm c} = \left(\frac{GM}{\Omega^2}\right)^{1/3}.
\end{equation}
Steady accretion requires $R_{\rm m}<R_{\rm c}$ so that matter can attach to
stellar field lines and reach the surface. When $R_{\rm m}>R_{\rm c}$ the
magnetosphere rotates faster than the disk and centrifugally expels the
inflow, this is the propeller regime
\citep{GhoshLamb1979b,Ghosh1995}. In our
\texttt{MESA} implementation, this condition for accretion is enforced using
\texttt{other\_adjust\_mdot} in \texttt{run\_star\_extras} as:
\begin{equation}
\text{\texttt{mstar\_dot}} =
\begin{cases}
\dot M_{\rm WD}, & R_{\rm m} < R_{\rm c}, \\
0,               & R_{\rm m} \ge R_{\rm c},
\end{cases}
\end{equation}
so accretion is switched off whenever the WD enters the propeller regime.

\subsection{Angular Momentum and External Torques}
\label{sec:angmom}

Having incorporated magnetic fields, rotation, and accretion into the WD structure, we now complete the model by adding the angular momentum evolution. In \texttt{MESA}, this is set by \textit{angular momentum conservation} during structural readjustment as the star contracts or expands \citep{Paxton13}. Under our assumption of solid body rotation and in the absence of external torques, the spin evolution follows as
\begin{equation}
\left(\frac{d\Omega}{dt}\right)_{\mathrm{amc}}
   = -\frac{2\Omega}{r}
     \left( \frac{\partial r}{\partial t} \right)_m
     \left( \frac{1}{2}\frac{d\ln i}{d\ln r} \right),
\label{eq:amc}
\end{equation}
where $i$ is the specific moment of inertia of a shell at mass coordinate $m$.  External torques enter through \texttt{extra\_omega\_dot} in the \texttt{other\_torque} routine, ensuring they are consistently added on top of the structural term such as,
\begin{equation}
\left(\frac{d\Omega}{dt}\right)_{\mathrm{total}}
   = \left(\frac{d\Omega}{dt}\right)_{\mathrm{amc}}
   + \left(\frac{d\Omega}{dt}\right)_{\mathrm{extra}}.
   \label{eq:allomega}
\end{equation}

\paragraph{Radiative torques.}
Magnetized, rotating WDs behave as oblique rotators when their magnetic and spin axes are misaligned, which is also consistent with the magnetic dichroism observed in several magnetic WDs \citep{Putney1992,Suto2023}. In such systems, the extended dipolar field produces electromagnetic (EM) radiation, while the magnetic deformation along with obliquity generates a time-varying quadrupole moment and hence CGW emission. These \textit{radiative torques} extract angular momentum from the star, driving the evolution of both $\Omega$ and $\chi$, given by \citep{CH1970,KMMB2020,das-mukhopadhyay,Spitkovsky2006,Das2025_MWD_CGW}
\begin{align}
\left(\frac{d\Omega}{dt}\right)_{\mathrm{rad}} &=- \frac{B_p^2 R_p^6 \Omega^3}{4c^3 I_{z^{\prime}z^{\prime}}} \left(1 + 1.1 \sin^2 \chi \right)
 \nonumber\\ 
& \hspace{-3.5em} -\frac{2G}{5c^5 I_{z^{\prime}z^{\prime}}} (I_{zz}-I_{xx})^2 \Omega^5 \sin^2 \chi \left(1 + 15 \sin^2 \chi \right),
\label{eq:dwdt}
\end{align}
\begin{align}
I_{z^{\prime}z^{\prime}} \frac{d\chi}{dt} &=- \frac{B_p^2 R_p^6 \Omega^2}{4c^3} \sin \chi \cos \chi
\nonumber\\
&- \frac{12G}{5c^5} (I_{zz}-I_{xx})^2 \Omega^4 \sin^3 \chi \cos \chi,
\label{eq:dxdt}
\end{align}
where $R_p$ is the polar radius and $I_{ij}$ are components of the moment of inertia tensor, with $I_{z^{\prime}z^{\prime}} = I_{zz}\cos^2\chi + I_{xx}\sin^2\chi$; $\hat{z}$ and $\hat{z}'$ denote the magnetic and spin axes, respectively. The radiative contribution to $d\Omega/dt$ is included through the \texttt{extra\_omega\_dot} and, as because \texttt{MESA} is 1D  and carries no information about $\chi$, we evolve $\chi$ as an auxiliary variable coupled to $\Omega$. Equations~\eqref{eq:dwdt} and~\eqref{eq:dxdt} are solved simultaneously using initial values of $I_{xx}$, $I_{zz}$, $B_p$, and $R_p$ obtained from the two-dimensional (2D) equilibrium WD models computed with stellar structure solver \texttt{XNS} (see Section~\ref{sec:gwmodel} for details) for given WD density, internal field and rotation.

\paragraph{Accretion torque.}
In addition, the disk-magnetosphere coupling exerts a torque $N_{\rm acc}$ on the WD. If the Keplerian disk at $R_{\rm m}$ rotates faster than the magnetosphere, the WD spins up; otherwise, the magnetic linkage brakes the star. The corresponding contribution to $d\Omega/dt$ (added through \texttt{extra\_omega\_dot}) is
\begin{equation}
\left(\frac{d\Omega}{dt}\right)_{\rm acc}
 = \frac{N_{\rm acc}}{I_{z'z'}}
 = \frac{n_{\rm s}\,\dot M_{\rm WD}\sqrt{GMR_{\rm m}}}{I_{z'z'}}.
 \label{eq:omgacc}
\end{equation}
Here $n_{\rm s}$ is the dimensionless torque function (of $R_{\rm m}/R_{\rm c}$) that parametrizes the disk-field coupling. In the classical GL model, the disk-field coupling is wide and extended well outside $R_{\rm c}$, so regions with $\Omega_{\rm K}<\Omega$ contribute strongly. As a result the net torque reverses sign already at $R_{\rm m}/R_{\rm c}\simeq0.5$, giving rise to three regimes: (i) $R_{\rm m}/R_{\rm c}\lesssim0.5$, the WD spins up as the torque per unit area is higher near the inner disk, (ii) $R_{\rm m}/R_{\rm c}\gtrsim0.5$, the braking region dominates, with $n_{\rm s}$ decreasing smoothly to zero and then becoming negative even while accretion continues and (iii) for $R_{\rm m}/R_{\rm c}>1$, a dead-disk with no accretion onto the WD, and spin-down as angular momentum is transported outward through the disk. In this work we use $n_{\rm s}(R_{\rm m}/R_{\rm c})$ from \citet{GhoshLamb1979b,Ghosh1995}, which spans approximately $+1.3$ to $ -0.2$.

Modern MHD simulations show that field-line reconnection severely restrict the coupling region \citep{Romanova2004}; \citet{DAngelo2010, DAngelo2012} -- DS model captures this behavior and gives torque reversal much later, near $R_{\rm m}/R_{\rm c}\approx0.9$, with the possibility of long-lived ``trapped'' states where the torque switches between spin-up and spin-down. GL predicts an earlier torque reversal and hence a lower final WD spin, while DS allows prolonged spin-up before the reversal, but if the system enters a trapped state for long a time it can experience substantial long-term spin-down.

The radiative term (Equation \ref{eq:dwdt}) and accretion term (Equation \ref{eq:omgacc}) are added into the extra  \texttt{extra\_omega\_dot} and thus Equation \eqref{eq:allomega} is solved for evolution of $\Omega$.

\subsection{Resulting spin evolution}
\label{sec:spincal}

We now solve Equations.~\eqref{eqdm}-\eqref{eqX}, \eqref{eq:Bdecay}, \eqref{eq:acc}, \eqref{eq:allomega}, \eqref{eq:dxdt} with \texttt{MESA} to follow the spin evolution of WDs in SSS. Table~\ref{tab:sssparams} lists the systems we target; as a fiducial case we evolve a CAL~83-like  WD of initial mass $M = 1.3\,M_\odot$ with accretion rate of
$\dot M_{\rm WD} = 10^{-7}\,M_\odot\,{\rm yr^{-1}}$, and initial spin $1/67\,{\rm Hz}$ -- as reported by \citet{Odendaal2014} with the X-ray pulsation study. The internal toroidal field is initialized with $B_m\sim10^{12}\,$G, and we assume a poloidal component with initial $B_p\approx B_m/10^{5\text{--}6}\sim10^{6}\text{--}10^{7}$\,G, consistent with the moderate surface fields ($\lesssim10^{7}$\,G) inferred for steadily accreting WDs in SSSs \citep{Kahabka2006}.

\begin{table}
\centering
\caption{Observed parameters (approximately) from \citet{Odendaal2014,Tomov1998,Zang2023,Prodhani2018,Greiner2000} for best studied and nearby SSS systems relevant for GW modelling.}
\setlength{\tabcolsep}{4pt}   
\renewcommand{\arraystretch}{1.05} 
\begin{tabular}{lccccc}
\hline
System 
& $M_{\rm WD}$ 
& $M_2$ 
& $\dot{M}_{\rm WD}$ 
& $\nu$ 
& $d$ \\[-2pt]

& ($M_\odot$) 
& ($M_\odot$) 
& ($M_\odot\,{\rm yr^{-1}}$) 
& (Hz) 
& (kpc) \\
\hline
CAL~83              & $\sim 1.3$ & 1.5--2.0 & $10^{-7}$ & $1/67$ & 50 \\
RX~J0019+2156     & $\sim 1.3$ & 2.0--3.0     & $10^{-7}$ & ---    & 2  \\
RX~J0925-4758    & $\sim 1.3$ & 1.0--2.0     & $10^{-7}$ & ---    & 10 \\
\hline
\end{tabular}
\label{tab:sssparams}
\end{table}

Three contributions shape the spin evolution:  
(i) the structural spin-up due to contraction during mass gain, (ii) the radiative spin-down from EM and GW torques, and (iii) the contribution from accretion torque, whose sign and magnitude depend on $R_{\rm m}/R_{\rm c}$. For the modest $B_p$ adopted here, the EM term is negligible and the GW term becomes relevant only as $\nu$ approaches the high end of the deciHz band; thus, throughout most of the SSS phase, the competition is primarily between contraction-driven spin-up and the accretion torque.

\begin{figure*}
\begin{center}
\includegraphics[scale=0.72]{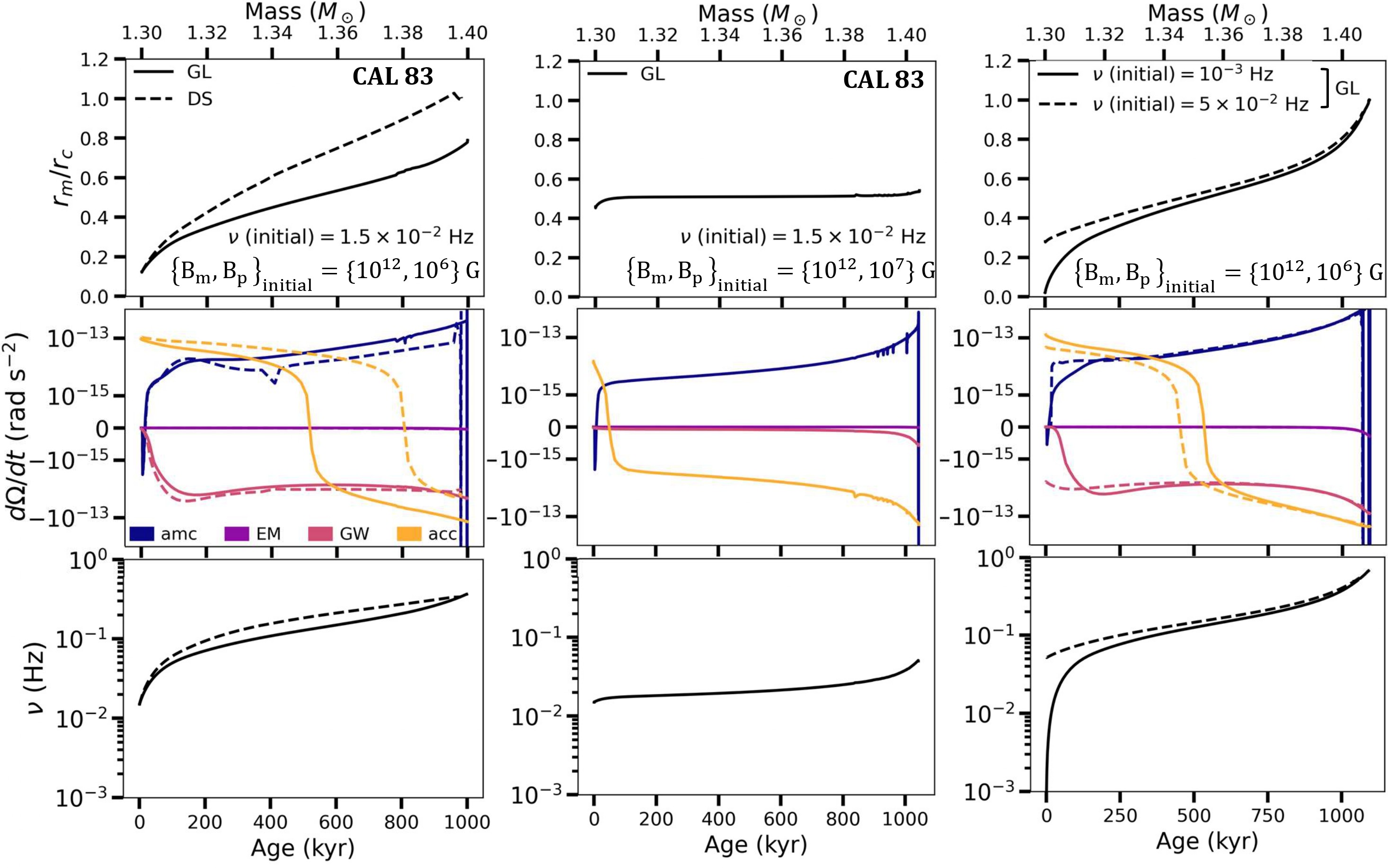}
\caption{Evolution of the magnetospheric to corotation radius ratio $R_{\rm m}/R_{\rm c}$ (first row), individual torque contributions to $d\Omega/dt$ (second row), and spin frequency $\nu$ (third row) for SSS models. The first column shows CAL~83 with $B_p=10^{6}$\,G using GL (solid) and DS (dashed) torques. The second column shows a higher-field CAL~83 sequence with $B_p=10^{7}$\,G using GL. The third column shows parameter studies for RX~J0019+2156 and RX~J0925$-$4758 with low ($10^{-3}$\,Hz; solid line) and high ($5\times10^{-2}$\,Hz; dashed line) initial spins using GL. In the second row, blue, purple, pink, and orange curves denote the structural, electromagnetic, gravitational wave, and accretion torques, respectively.}
\label{fig:nuevolcal}
\end{center}
\end{figure*}

These behaviors are visualized in Figure \ref{fig:nuevolcal}: the top panels show the evolution of $R_{\rm m}/R_{\rm c}$, the middle panels show the individual torque contributions to $d\Omega/dt$, and the bottom panels show the resulting spin evolution beginning from CAL~83’s observed spin at present $\nu=1.5\times 10^{-2}$ Hz. The \textit{first column} shows the evolution for a WD with a moderate initial surface field $B_p\sim 10^{6}$ G using both GL and DS torques. Initially $R_{\rm m}/R_{\rm c}<0.5$, thus the WD spins up due to accretion torque and $R_{\rm m}/R_{\rm c}$ grows as spin and field increase during steady accretion. Accretion torque reversal then occurs at $R_{\rm m}/R_{\rm c}\simeq 0.5$ for GL and $0.9$ for DS, leading in the GL case to an earlier shift from spin-up to either spin-down due to accretion torque only. However, depending on the balance between structural and accretion terms there is a transition from faster spin up to slower spin up. 
Once $R_{\rm m}/R_{\rm c} > 1$, accretion is inhibited and the WD spins down due to angular momentum transport from the star to the dead disk, and also GW, EM torques. This spin-down reduces $R_{\rm m}/R_{\rm c}$, allowing accretion to resume, which in turn drives renewed spin-up. The system, therefore, undergoes cyclic transitions between accreting and non-accreting phases as it approaches the Chandrasekhar mass.
\footnote{About 15-30\% of $M_2$ is expected to be supplied during the thermal mass transfer, providing $\sim0.5 M_\odot$ of transferred material. With further wind-regulated accretion, an initially massive CAL 83 like WD can still gain enough mass ($\sim 0.15 M_\odot$) to reach Chandrasekhar mass within the thermal transfer phase potentially leading to a Type~Ia supernova \citep{Hachisu1999}.} 

The DS torque produces a stronger initial spin-up due to its larger positive accretion torque, but this is progressively offset by enhanced centrifugal support, which reduces the rate of stellar contraction and weakens the contraction-driven structural spin-up (see Appendix, Figure \ref{fig:append}). When the accretion torque becomes negative, the structural term dominates, causing the overall spin-up in the DS case to flatten in the later phase. In contrast, the GL model shows slower spin-up at early times but maintains a relatively stronger increase at later stages when the contraction-driven structural spin-up is stronger than the DS model. As a result, DS exhibits faster early spin-up followed by a plateau, while GL shows delayed but sustained spin-up, with both models converging to a similar final spin. Thus, while the evolutionary paths differ, the terminal spin is largely insensitive to the torque prescription.

The DS model was originally developed to describe weak, channeled, or partially inhibited accretion. However, CAL~83 and SSSs in general are characterized by sustained, steady accretion. Thus,  we adopt the GL torque as our default for the remaining of this work. However, we cannot meaningfully assess this choice observationally, as there is no measurement or useful limit on the spin frequency or its derivative.

Since the surface magnetic field of CAL~83 is not observationally constrained, in the \textit{second column} we explore a higher-field sequence with initial $B_p \sim 10^{7}$ G. The larger magnetosphere places $R_{\rm m}/R_{\rm c}$ close to the torque flip threshold ($\simeq 0.5$) from the outset, limiting the efficiency of spin-up and resulting in only a slight increase in the final spin relative to the initial value. For $B_p \gtrsim 10^{7}$ G, $R_{\rm m}/R_{\rm c}$ would remain $>1$ from the beginning, suppressing accretion entirely, which is incompatible with the observed steady accretion in SSSs. We, therefore, take $B_p \lesssim 10^{7}$ G as a conservative upper limit. Accordingly, we adopt $B_p \approx 10^{6}$ G for the remainder of this work, as such moderate fields are better representative of the SSS population than the rare high field cases.
 
In the \textit{third column}, we examine RX~J0019+2156 and RX~J0925$-$4758, which, as Galactic SSSs, would produce higher CGW strains due to their closer distances, if their frequencies are similar to CAL 83. However, there is a lack of knowledge for their measured spin frequencies. We, therefore, perform a parameter study with low ($10^{-3}$ Hz) and high ($5\times10^{-2}$ Hz) initial spins. The final spins converge to similar values despite the different starting points: models with higher initial $\nu$ enter the accretion spin-down regime earlier owing to their larger initial $R_{\rm m}/R_{\rm c}$, while low-spin models experience a prolonged spin-up phase. Remarkably, the final spins in all cases are close to CAL~83’s present observed value, implying that comparable SSSs naturally converge to similar frequencies due to sufficient angular momentum transport from the disk, once steady accretion is established.

\section{gravitational wave emission}
\label{sec:gwmodel}

Having obtained the time-dependent WD structure, magnetic field and spin evolution from the \texttt{MESA} framework described in Section~\ref{sec:wdmodel}, we now compute the resulting quadrupolar deformation essential for GW emission. Because magnetic stresses break spherical symmetry, deformation or ellipticity cannot be extracted from the 1D \texttt{MESA} model. They require a full 2D magneto-hydrostatic stellar structure solver, e.g. \texttt{XNS 4.0} \citep{PBDZ2014,sold2021main}, to calculate the principal moments of inertia along different axes. For each snapshot along the \texttt{MESA} evolutionary track, the instantaneous $[M,R,B_m,\Omega]$ are used as the reference to \texttt{XNS} to compute a corresponding 2D equilibrium model, from which we extract $I_{xx}$, $I_{zz}$ and, hence, $\epsilon = |I_{zz}-I_{xx}|/I_{xx}$ for axisymmetric models.

\subsection{Magnetized equilibrium from \textnormal{\texttt{XNS}}}
\label{subsec:xns}

\begin{figure}[t]
\begin{center}
\includegraphics[width=\columnwidth]{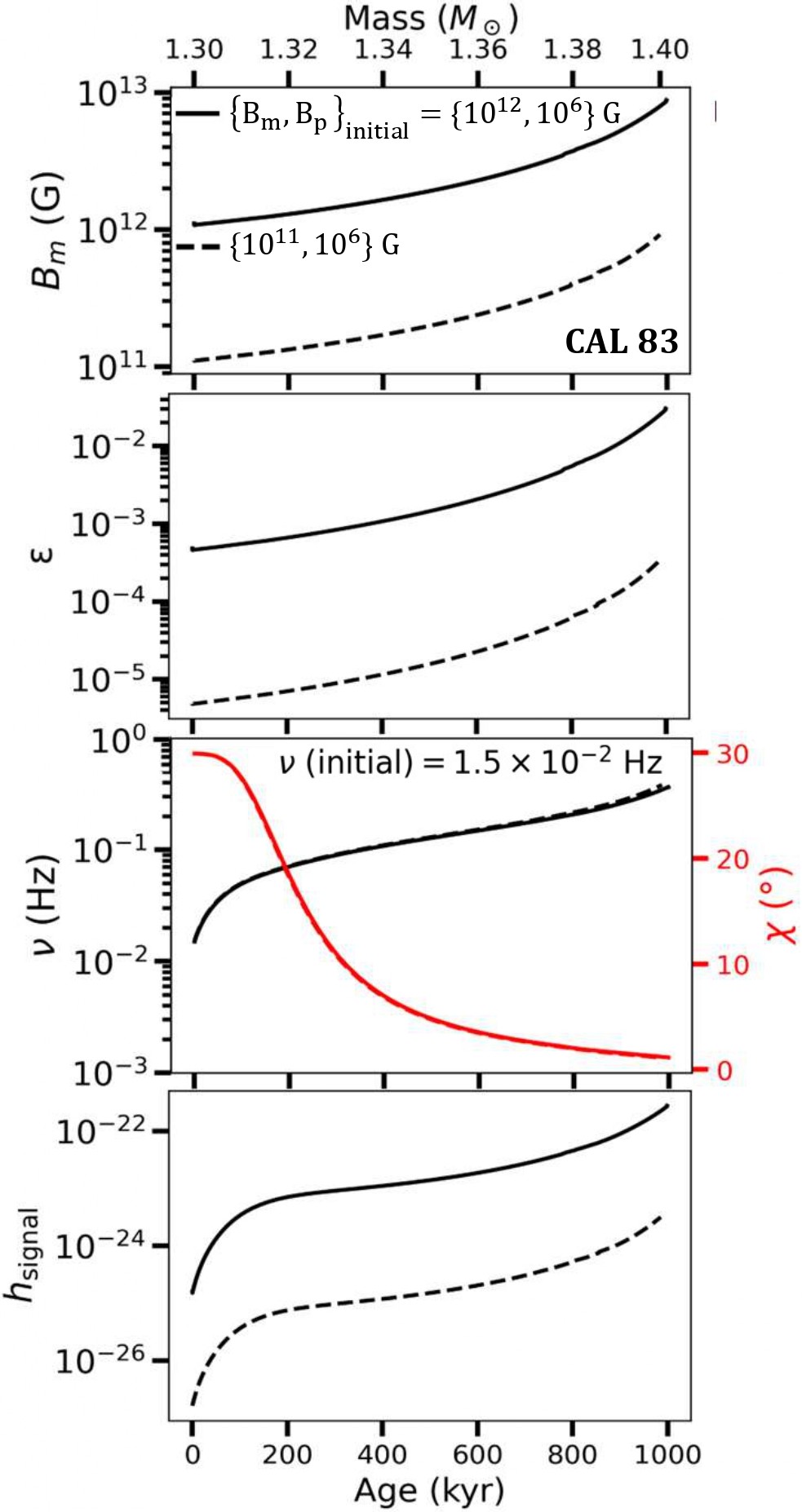}
\caption{Evolution of the internal magnetic field $B_m$ (first panel), ellipticity $\epsilon$ (second panel), spin frequency $\nu$ (black) and obliquity angle $\chi$ (red; third panel), and characteristic GW strain $h_{\rm signal}$ (fourth panel) for CAL~83. Solid curves correspond to the fiducial model with initial magnetic field profile having maximum internal magnitude $B_m = 10^{12}$\,G while dashed curves show a comparison model with initial $B_m = 10^{11}$\,G.}
\label{fig:spinchi}
\end{center}
\end{figure}

\begin{figure*}[t]
\begin{center}
\includegraphics[scale=0.7]{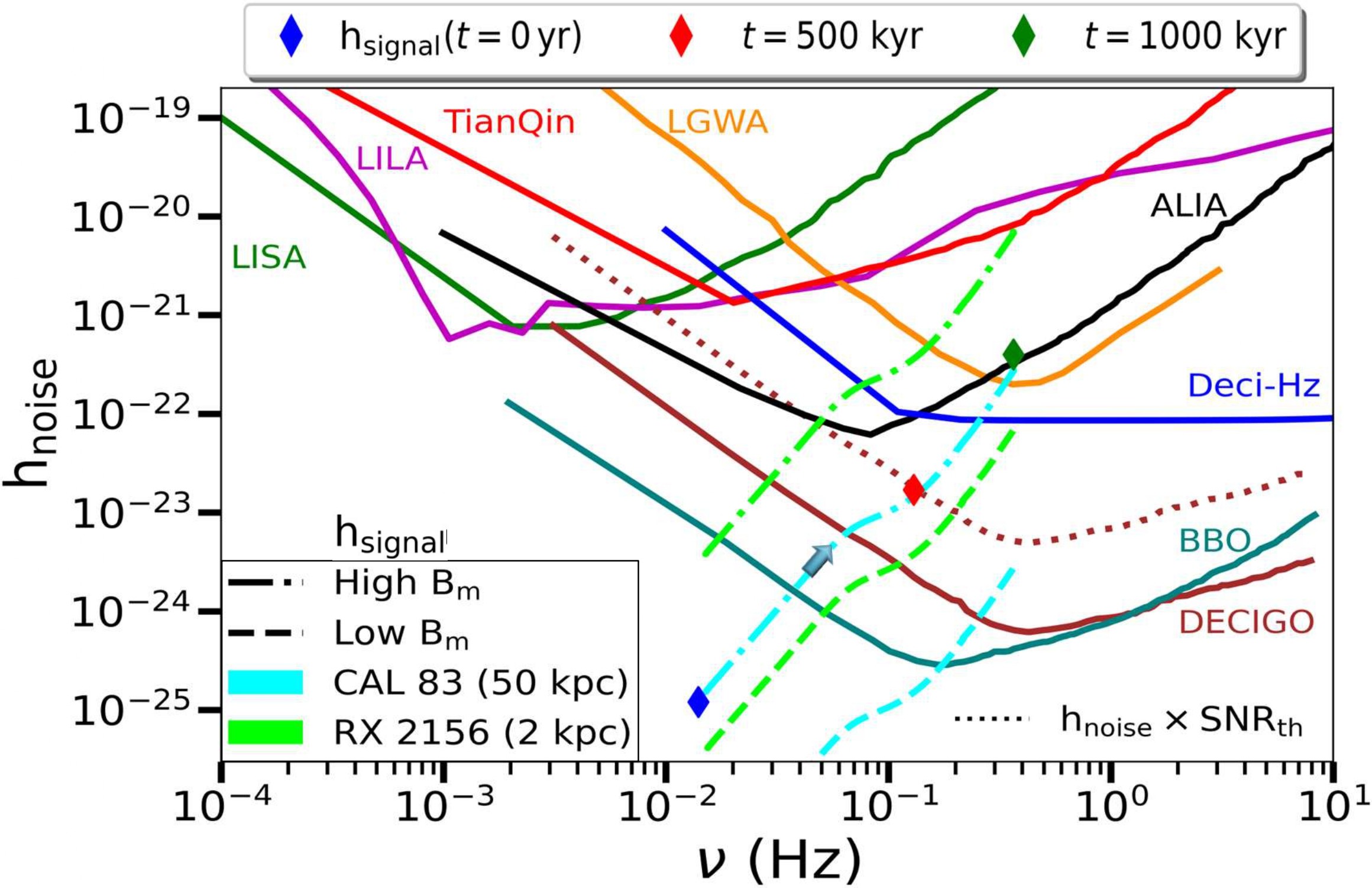}
\caption{Characteristic GW strain $h_{\rm signal}$ for CAL~83 and RX~J0019+2156 (or RX 2156), compared with the noise of milli-deciHz GW detectors. Dash-dotted and dashed curves correspond to models with initial internal magnetic fields $B_m=10^{12}$\,G (high) and $10^{11}$\,G (low), respectively. Diamond markers indicate $h_{\rm signal}$ for CAL 83 (with high $B_m$ model) at $t=0$ (blue), 500\,kyr (red), and 1000\,kyr (green) along the evolutionary track. Sky-blue curves show CAL~83 residing at $d=50$ kpc, while green curves show a scaled version at 2\,kpc representative of RX~J0019+2156. Dotted curves indicate ${\rm SNR}_{\rm th}\times h_{\rm noise}$ for DECIGO.}
\label{fig:sensboth}
\end{center}
\end{figure*}

\texttt{XNS} solves the coupled Einstein-Maxwell equations for stationary, axisymmetric, rotating, magnetized stars in ideal MHD. The stress--energy tensor is
\begin{equation}
T^{\mu\nu}
  = (e + P + b^2)\,u^\mu u^\nu
    - b^\mu b^\nu
    + \left(P + \frac{1}{2} b^2\right) g^{\mu\nu},
\label{eq:tmunu}
\end{equation}
where $e$ is the energy density, $P$ is the fluid pressure, $u^\mu$ is the four-velocity (whose azimuthal component encodes rotation),
$b^\mu \equiv {}^{\ast}F^{\mu\nu}u_\nu$ is the magnetic field measured in the comoving (fluid) frame, $F^{\mu\nu}$ is the electromagnetic field tensor and ${}^{\ast}F^{\mu\nu}$ is its dual, $B^\mu$ is the stellar magnetic field whose spatial components are $B^r$, $B^\theta$, and $B^\phi$, and $g^{\mu\nu}$ is the spacetime metric. The stellar matter satisfies magneto-hydrostatic equilibrium condition
\begin{equation}
\nabla_\nu T^{\mu\nu} = 0,
\label{eq:hydroeq}
\end{equation}
together with Maxwell’s equations,
\begin{equation}
\nabla_\nu F^{\mu\nu} = 4\pi J^\mu,
\qquad
\nabla_\nu {}^{\ast}F^{\mu\nu} = 0,
\label{eq:maxwell}
\end{equation}
where $J^\mu$ is the four-current. Closure is provided by a cold WD equation of state, which we approximate with a relativistic polytrope $P = K\rho^{4/3}$, where $K = 4.8 \times 10^{14}$ in CGS units for degenerate electron gas. This polytropic EOS differs from the temperature dependent \texttt{Skye} EOS used in \texttt{MESA}, but \texttt{XNS} neglects thermal effects by construction the temperature effect is negligible for the WDs (i.e. $\rho_c$) under consideration.

For each chosen combination of $\rho_c$, internal magnetic field configuration (we use purely toroidal fields), $B_{m}$, $\Omega$ (assumed uniform throughout the star), \texttt{XNS} returns the full 2D equilibrium, including $M$, circumferential radius $R$, $I_{xx}$, $I_{zz}$ and magnetic field profile $B(r,\theta)$. In this work, we compute a grid of WD equilibria with \texttt{XNS} spanning the relevant range of $(\rho_c, B_{\rm max}, \Omega)$ as found out from the WD's evolutionary track with \texttt{MESA}. At each time step we thus obtain $\epsilon$ consistent with the instantaneous mass, radius, internal field, and rotation rate results from Secion ~\ref{sec:wdmodel}.

\subsection{GW strain from a deformed, oblique WD}
\label{subsec:gw_strain}

If the symmetry axis of the magnetic deformation is misaligned with the spin axis, the star emits CGWs with intrinsic GW amplitude \citep{BG1996}
\begin{equation}
h_0 = \frac{2G}{c^4}\,
      \frac{4\pi^2 \nu^2\,\epsilon\,I_{xx}}{d},
\label{eq:h0}
\end{equation}
where $d$ is the source distance. The observed GW strain depends on $\chi$ and the inclination $i$ of the spin axis to the line of sight as:
\begin{equation}
h = f(\chi,i)\,h_0,
\label{eq:h_obs}
\end{equation}
where $f(\chi,i)$ encodes the relative power in the GW components at $\nu$ and $2\nu$. Following DMB2025,
\begin{align}
f(\chi,i)
 &= \frac{\sin\chi}{2\sqrt{2}}
 \Big[(1+\cos^2 i)^2 \sin^2\chi
      + 4\cos^2 i \cos^2\chi\Big]^{1/2}
 \nonumber\\
 &\quad \times (2\cos^2\chi - \sin^2\chi).
\label{eq:fchi}
\end{align}
As \texttt{XNS} computes stationary, axisymmetric equilibria, after extracting $\epsilon$ we assume a value of $\chi$ during post-processing to calculate the GW strain.
When $\chi$ and $i$ are unknown, we adopt the orientation-averaged value $\langle f\rangle_{i,\chi} \simeq 0.147$ \citep{Das2025_MWD_CGW}.

Coherent integration over observation time $T_{\rm obs}=4$ yr stacks $\nu T_{\rm obs}$ number of cycles and enhances the effective signal amplitude of source to the characteristic signal
\begin{equation}
h_{\rm signal}
  = h_0 \sqrt{\nu T_{\rm obs}},
\label{eq:h_signal}
\end{equation}
which must exceed the detector noise amplitude $h_{\rm noise}(\nu)$ to be detected by the detector. Thus, a signal is detectable only if the signal-to-noise ratio (SNR) 
\begin{equation}
    \mathrm{SNR} = \frac{h_{\rm signal}}{h_{\rm noise}},
\end{equation}
exceeds the threshold $\mathrm{SNR}_{\rm th} \approx 8$ for more than $\sim$95\% detection efficiency \citep{Tang2024}.

To illustrate how the CGW signal from CAL~83 builds up during accretion, in Figure~\ref{fig:spinchi} we track the coupled evolution of $B_m$, $\epsilon$, $\nu$, $\chi$, and the resulting strain $h_{\rm signal}$. The solid curves correspond to our fiducial CAL~83 model with initial $B_m = 10^{12}$\,G (identical to the GL case shown in the first column of Figure~\ref{fig:nuevolcal}), while dashed curves show a comparison sequence with a weaker initial field $B_m = 10^{11}$\,G.

The first panel shows the amplification of $B_m$ driven by stellar contraction during mass growth. The second panel shows the corresponding increase in $\epsilon$, reflecting the increasing magnetic deformation. The third panel shows the evolution of  $\nu$ (black) together with $\chi$ (red), evolved self-consistently using Equations ~\eqref{eq:dxdt} and \eqref{eq:allomega} with an adopted initial $\chi = 30^\circ$. As accretion proceeds, $\nu$ increases while $\chi$ gradually aligns due to combined EM and GW torques. The fourth panel shows the resulting $h_{\rm signal}$.

In Figure~\ref{fig:sensboth} we evaluate the detectability of SSSs by comparing $h_{\rm signal}$ computed for a coherent observation time $T_{\rm obs}=4$ yr, with the noise curves of various milli--deciHz GW detectors. It shows $h_{\rm signal}$ for the fiducial $B_m = 10^{12}$\,G model at three representative epochs ($t=0$, 500, and 1000\,kyr), together with the detector sensitivities. For CAL~83, at a distance of 50\,kpc (sky-blue dash-dotted curve), the signal remains below the LISA sensitivity but exceeds the detection thresholds of other next generation detectors such as Deci-Hz, DECIGO, BBO, ALIA, and LGWA for a 4 yr coherent observation once $h_{\rm signal} > {\rm SNR}_{\rm th}\times h_{\rm noise}$. For DECIGO, we explicitly show the ${\rm SNR}_{\rm th}\times h_{\rm noise}$ curve to illustrate the detectability.

The same figure also shows a scaled version of the CAL~83 track at a distance of 2\,kpc (green dash-dotted curve), representative of the nearby Galactic SSS RX~J0019+2156. Although the spin of RX~J0019+2156 is unknown, our spin-evolutions (see Figure~\ref{fig:nuevolcal}) demonstrate that SSSs converge to CAL~83 like frequencies once steady accretion is established, largely independent of their initial rotation. This scaling, therefore, provides a realistic estimate of its CGW detectability.

For completeness, we also include the corresponding tracks for the weaker-field case $B_m = 10^{11}$ G (dashed curves) for CAL~83 and RX~J0019+2156. Although these models produce systematically lower strains, they still approach or exceed the sensitivities of DECIGO and BBO. Since RX~J0925$-$4758 lies at an intermediate distance ($\sim10$ kpc), its expected detectability follows straightforwardly from these bounds. This demonstrates that CGW detectability from SSSs is robust across a wide and observationally unconstrained range of internal magnetic field strengths. Steady accretion simultaneously increases $B_m$, $\epsilon$, and $\nu$, causing $h_{\rm signal}$ to grow with time and making SSSs compelling CGW sources. For known SSSs, such as CAL~83 and RX~J0019+2156, where the sky position and distance are constrained, a CGW detection could probe the WD’s internal magnetic field and rotation.

\subsection{Signal to noise ratio}
\label{snr}

We now Calculate cumulative SNR to asses detectability over $T_{\mathrm{obs}}$ with $\mathcal{N}$ number of shorter time stacks $T_{\text{s}}=T_{\text{obs}}/\mathcal{N}$ for two frequencies of GW emission is  \citep{Jaranowski1998, MAG2008}
\begin{equation}
\langle \text{SNR} \rangle = \sqrt{ \langle \text{SNR}_\nu^2 \rangle + \langle \text{SNR}_{2 \nu}^2 \rangle },
\label{eq:snr}
\end{equation}
where
\begin{equation}
\langle \text{SNR}_\nu^2 \rangle =  \frac{\sin^2 \zeta}{100} \, \sum_{s=1}^{\mathcal{N}}\frac{ T_{\text{s}} \, h_0^2\, \sin^2 2\chi}{S_n(\nu)},
\label{eq:snrw}
\end{equation}
\begin{equation}
\langle \text{SNR}_{2 \nu}^2 \rangle = \frac{4 \,\sin^2 \zeta}{25} \, \sum_{s=1}^{\mathcal{N}} \frac{T_{\text{s}} \, h_0^2 \, \sin^4 \chi}{S_n(2\nu)}.
\label{eq:snr2w}
\end{equation}
Here, $\zeta$ is the angle between the interferometer’s arms: $\zeta = 60^\circ$ for space-based detectors and $90^\circ$ for ground based ones. The PSD data $S_n(\nu)$ are collected from publicly available sensitivity curves (e.g., \citealt{Moore2015}:
\url{https://gwplotter.com/}).

\begin{figure}
\begin{center}
\includegraphics[width=\columnwidth]{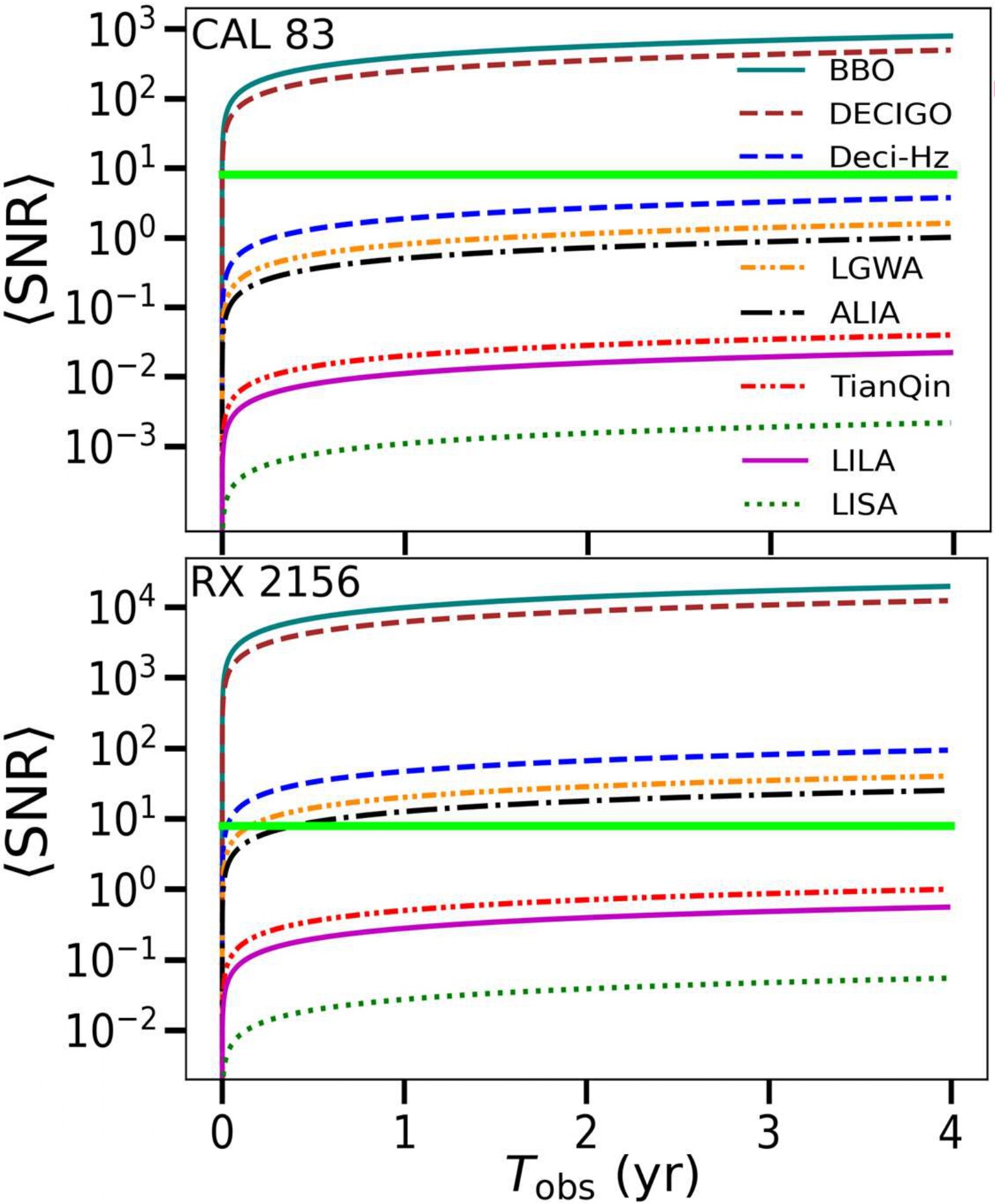}
\caption{Cumulative signal-to-noise ratio (SNR) as a function of observation time $T_{\rm obs}$ for CGW emission from CAL~83 (top panel) and RX~J0019+2156 (bottom panel), assuming the fiducial internal magnetic field $B_m=10^{12}$\,G. The horizontal green line marks the detection threshold ${\rm SNR}_{\rm th}=8$. Curves correspond to different GW detectors as labeled.}
\label{fig:snr}
\end{center}
\end{figure}

In Figure~\ref{fig:snr}, we show the cumulative SNR for a continuous 4 yr observation of CAL~83 near the end of its accretion phase, when the CGW signal is strongest (See fourth panel of Figure \ref{fig:spinchi}). For the fiducial $B_m=10^{12}$\,G model, CAL~83 would be readily detectable by DECIGO and BBO. The lower panel shows the corresponding SNR for RX~J0019+2156, scaled to its Galactic distance (2\,kpc). Owing to its proximity, the source achieves significantly higher SNRs and would be detectable by DECIGO, BBO, Deci-Hz, ALIA, and LGWA mission well within the 4 yr mission lifetime. Because the signal peaks near the end of the accretion phase and close to the Chandrasekhar mass, a CGW detection could identify a candidate Type~Ia progenitor as accreting WDs, prior to the appearance of the electromagnetic supernova explosion signature.

\subsection{Detection probability: Number of WDs in SSSs}
\label{Detection plausibility}

All such WDs should be detectable out to a maximum distance $d$ where their signal matches the threshold detector sensitivity of DECIGO for 4 yr mission:
\begin{align}
    h_{\text{noise}}(\nu) &= \frac{h_{\text{signal}}}{\text{SNR}_{\text{th}}} = f(\chi)\,\frac{2G}{c^4}\,\frac{4\pi^2\nu^2\,\epsilon\,I_{xx}}{d} \frac{\sqrt{\nu T_{\mathrm{obs}}}}{\text{SNR}_{\text{th}}},
    \label{eq:h_detector}
\end{align}
where $f(\chi)=\langle f(\chi,i) \rangle_i$, the averaged function over all possible inclination, as $i$ is not known observationally. Thus, $d$ can be calculated for each $\nu$, $\chi$ with $h_{\text{noise}}$ of DECIGO and $\epsilon$ from the simulated WD model, i.e.,
\begin{align}
    d(\nu,\chi) &= f(\chi)\frac{2G}{c^4} \frac{4\pi^2\nu^2 \epsilon I_{xx}}{h_{\text{noise}}(\nu)}\frac{\sqrt{\nu T_{\mathrm{obs}}}}{\text{SNR}_{\text{th}}}.
    \label{eq:d_fchi}
\end{align}

\begin{figure}[t]
\begin{center}
\includegraphics[width=0.9\columnwidth]{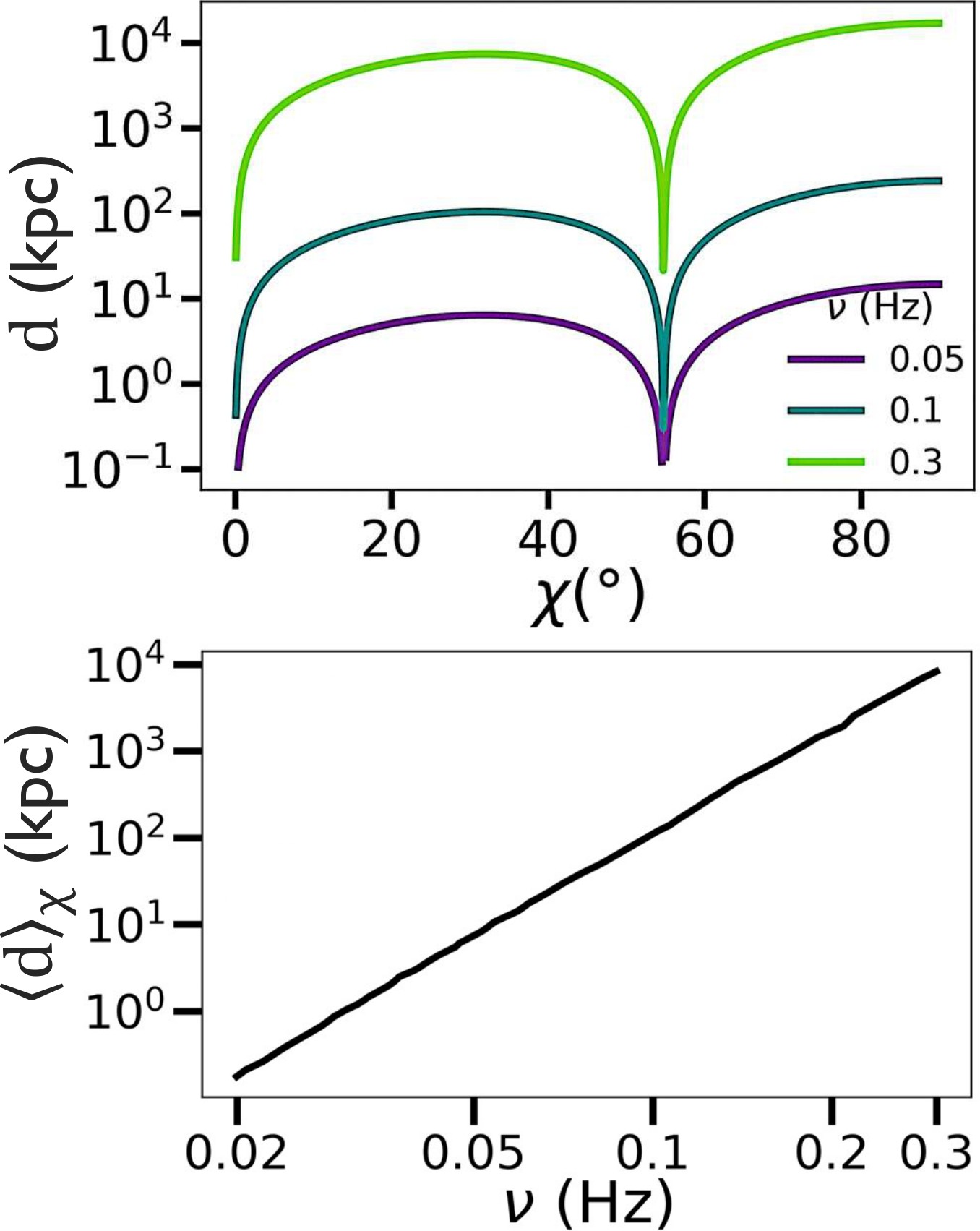}
\caption{Maximum detectable distance $d$ for CGW emission from SSS WDs with DECIGO for a 4 yr observation. Top panel: $d$ as a function of obliquity angle $\chi$ for representative spin frequencies $\nu=0.05$, 0.1, and 0.3\,Hz, computed using the ellipticity of the fiducial CAL~83 model with $B_m=10^{12}$\,G. Bottom panel: orientation-averaged detection distance $\langle d\rangle_\chi$ as a function of spin frequency $\nu$.}
\label{fig:3ddecigo}
\end{center}
\end{figure}

The top panel of Figure~\ref{fig:3ddecigo} shows the maximum distance out to which DECIGO can detect SSS WDs as a function of $\chi$ for representative spin frequencies, using the ellipticity $\epsilon = 5\times10^{-4}$ of the fiducial CAL~83 model with $B_m = 10^{12}$ G. This internal field strength need not be primordial, but can also be reached through magnetic amplification during accretion (see the first column of Figure~\ref{fig:spinchi}). Since the obliquity is not observationally constrained for SSSs, the bottom panel shows the corresponding orientation-averaged distance $\langle d\rangle_\chi$. As demonstrated by the spin evolution models in Figure~\ref{fig:nuevolcal}, WDs undergoing steady accretion generically reach spin frequencies $\nu \sim 0.1$ Hz on relatively short timescales for moderate surface fields, making this frequency range representative of the SSS population.

At $\nu \sim 0.1$ Hz, DECIGO can probe SSS WDs out to distances of $\sim 80-90$kpc for the adopted ellipticity here, encompassing the entire MW and LMC. Population synthesis studies predict $\gtrsim 1000$ SSSs in the Galaxy and a few hundred in the LMC \citep{Rappaport1994}, while only $\sim 1-10\%$ are currently observed and others are obscured due to interstellar absorption \citep{Greiner2000}. DECIGO’s reach, therefore, implies sensitivity to the bulk of this obscured population. Even for more slowly rotating systems with $\nu \sim 0.05$ Hz, DECIGO can probe up to distances of $\sim 8$ kpc. Scaling by the accessible Galactic volume, this corresponds to sensitivity to $\sim 1000 \times (8/15)^2 \simeq 280$ Galactic SSSs, demonstrating that CGW observations provide a powerful and complementary probe of the hidden SSS WD population.

\subsection{Distinguishing GWs from Other Sources}
\label{gwdist}
In order to establish that a detected CGW signal originates from a WD in an SSS, it is necessary to distinguish our sources from alternative emitters in the same frequency band. In the milli Hz range, persistent GW signals are also produced by AM\,CVn systems (ultracompact interacting binaries) and detached double-WD binaries (DWDs), which dominate the Galactic population \citep{Lamberts2019}. These systems can appear observationally similar to CGWs from accreting WDs. The key discriminant is the ability to resolve the frequency derivative $\dot{\nu}$ in long duration observations \citep{Takahashi2002}, which separates SSS accretors from the binary foreground.

AM\,CVn-s typically have orbital frequency derivatives of $\lvert\dot{\nu}\rvert \sim 10^{-17}$--$10^{-16}$\,Hz\,s$^{-1}$ \citep{Gokhale2007}, while DWDs can reach $\lvert\dot{\nu}\rvert \sim 10^{-13}$\,Hz\,s$^{-1}$ depending on their orbital frequency \citep{Peters1964,Maggiore2008}. Our model predicts $\lvert\dot{\nu}\rvert \sim 4\times10^{-16}$--$10^{-13}$\,Hz\,s$^{-1}$ (calculated from $\nu$ evolution in Figure \ref{fig:spinchi}) depending on the evolutionary stage of the SSS and the initial WD spin before accretion. $\lvert\dot{\nu}\rvert$ lie above DECIGO’s detectable resolution; $\Delta\dot{\nu} \approx 3.4\times10^{-17}$\,Hz\,s$^{-1}$ (calculated from \citealt{Takahashi2002}) for a 4 yr observation with $\mathrm{SNR}=8$. The range of $\lvert\dot{\nu}\rvert$ suggests clear distinction of SSS WDs from AM\,CVn-s, though not necessarily from DWDs. However, at frequencies $\nu > 0.02$\,Hz the DWD population becomes essentially absent \citep{Lamberts2019}. A CGW in this frequency band is reachable for SSS WDs as soon as accretion starts (as shown in Figure \ref{fig:spinchi}), would therefore almost certainly originate from a WD in an SSS.

However, isolated magnetized WDs can also exhibit CGWs with $|\dot{\nu}|\sim10^{-15}$--$10^{-14}\,\mathrm{Hz\,s^{-1}}$ due to spin-down (i.e., $\dot{\nu}<0$), as shown in our earlier isolated WD study, (DMB2025). In contrast, WDs in SSSs generally spin up ($\dot{\nu}>0$), which provide a robust discriminant between SSS accretors and isolated WDs as CGW sources.

\section{Conclusion}
\label{sec:conc}

We have demonstrated that SSSs constitute a robust and physically motivated class of CGW sources, arising naturally from the coupled evolution of mass accretion, magnetic field amplification, and rotational spin-up in WDs. By consistently linking time-dependent WD evolution with \texttt{MESA} to magnetized equilibria obtained from the Einstein-Maxwell solver \texttt{XNS}, we have quantified the magnetic deformation and CGW emission throughout the SSS phase.

Our results show that steady accretion drives internal toroidal fields to $B_m \sim 10^{12}-10^{13}$\,G and induces ellipticities $\epsilon \sim 10^{-6}-10^{-2}$ in massive WDs. At the same time, disk-field coupling and contraction-driven spin-up lead toward spin frequencies $\nu \sim 0.05-0.3$\,Hz, depending on the initial dipolar magnetic field, but largely independent of the initial rotation.

The resulting CGW from well-studied SSSs such as CAL~83 in LMC grows during the accretion phase and becomes detectable by planned detectors such as DECIGO, BBO for multi-yr coherent observations, while remaining below LISA sensitivity. Nearby galactic SSSs such as RX~J0019+2156 emerge as higher SNR CGW sources, thus can be also detectable by Deci-Hz, ALIA, LGWA. Importantly, detectability remains robust across two-three order of magnitude uncertainty in the internal magnetic field strength, underscoring that CGW emission is a generic outcome of accretion in SSSs rather than a fine tuned scenario.

CGW observations further provide a powerful probe of the obscured SSS population. Depending on the ellipticities and spins of WDs in SSSs, DECIGO can access the SSS population in our Galaxy and Magellanic Clouds, implying sensitivity to hundreds to thousands of systems that remain undetected in soft X-rays. The sign and magnitude of $\dot{\nu}$ offer a clear discriminator against the dominant milli Hz foreground from detached DWDs and AM\,CVn-s, and also distinguish accreting SSS WDs from isolated WD CGW emitters.

Finally, a targeted CGW detection from known SSSs such as CAL~83, RX~J0019+2156, RX~J0925-4758 would directly probe the WD’s internal magnetic field and rotation. CGW detection from a near-Chandrasekhar SSS could also identify the system as a potential pre-explosion Type~Ia progenitor candidate. Our results establish SSSs as prime CGW targets for upcoming deciHz missions and motivate dedicated targeted and blind searches as a new window onto rapidly rotating, magnetized, accreting WDs in the local Universe.

\section*{Acknowledgments} MD thank Nils Andersson for suggestions about magnetic field, David Bour for discussion about MESA, and Soumallya Mitra for insightful discussions on computational implementation. She also acknowledges the Prime Minister’s Research Fellowship (PMRF) scheme, with Ref. No. TF/PMRF-22-5442.03. TB acknowledges support  of the  Polish National Science Center (NCN) grant 2023/49/B/ST9/02777 and grant Maestro (2018/30/A/ST9/00050). BM acknowledges a project funded by SERB, India, with Ref. No. CRG/2022/003460, for partial support towards this research.

\section*{Appendix}

\begin{figure*}
\begin{center}
\includegraphics[scale=0.82]{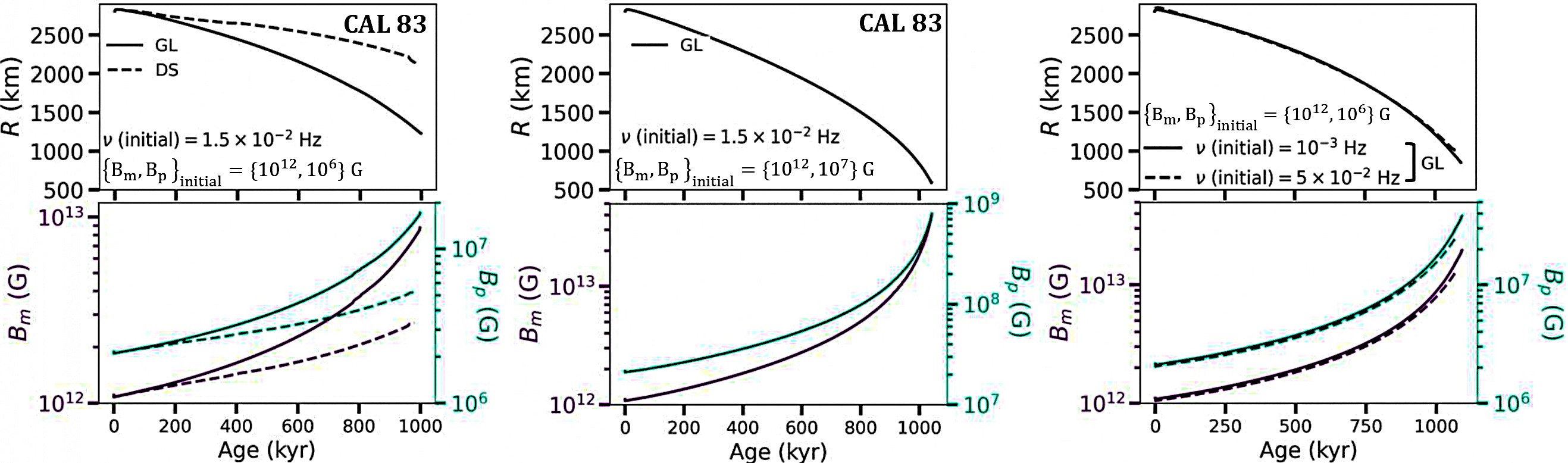}
\caption{
Evolution of the WD radius (top panels), and internal toroidal ($B_m$) and surface
poloidal ($B_p$) magnetic fields (bottom panels) for accreting WD
models. The columns correspond to the same models shown in
Figure~\ref{fig:nuevolcal}.
}
\label{fig:append}
\end{center}
\end{figure*}

Figure~\ref{fig:append} illustrates the evolution of the WD radius (top row)
and the resulting magnetic field amplification (bottom row) due to radial contraction for the
models discussed in Section~\ref{sec:spincal}. The first column shows that the DS
torque leads to enhanced centrifugal support, which slows stellar contraction
relative to the GL case and weakens the contraction-driven structural spin-up.
This effect explains the late-time flattening of the DS spin evolution when the accretion torque becomes negative and contraction-driven structural spin-up is the only dominant part, as seen in
Figure~\ref{fig:nuevolcal}. The remaining columns correspond directly to the other columns shown in Figure~\ref{fig:nuevolcal}.

\bibliographystyle{aasjournal}
\bibliography{draft_m3, reference} 

\end{document}